\documentclass{article}

\usepackage{arxiv}

\usepackage[utf8]{inputenc} 
\usepackage[T1]{fontenc}    
\usepackage{hyperref}       
\usepackage{url}            
\usepackage{booktabs}       
\usepackage{amsfonts}       
\usepackage{nicefrac}       
\usepackage{microtype}      
\usepackage{lipsum}

\usepackage{mathtools} 
\usepackage{graphicx,setspace}

\title{A Landauer Formula for Bioelectronic Applications}

\author{ Eszter Papp, Dávid P. Jelenfi, M\'at\'e T. Veszeli and G\'abor Vattay\\
Department of Physics of Complex Systems, \\
E{\"o}tv{\"o}s Lor\'and University \\
H-1117 Budapest, P\'azm\'any P\'eter s\'et\'any 1/A, Hungary\\
\texttt{vattay@elte.hu}\\
}

\begin{document}
\maketitle

\begin{abstract}

Recent electronic transport experiments using metallic contacts attached to proteins identified some 'stylized facts' which
contradict conventional wisdom that increasing either the spatial distance between the electrodes or the temperature suppresses conductance exponentially.
These include nearly temperature independent conductance over the protein in the 30-300K range, distance independent conductance within a single-protein in the 1-10 nm range and an anomalously large conductance in the 0.1-10 nS range. 
In this paper we develop a generalization of the low temperature Landauer formula which can account for the joint effects of tunneling and decoherence and can explain these new experimental findings. We use novel approximations which greatly simplify the mathematical treatment and allow us to calculate the conductance in terms of a handful macroscopic parameters instead of the myriads of microscopic parameters describing the details of an atomic level quantum chemical computation. The new approach makes it possible to get 
predictions for the outcomes of new experiments without relying solely on high performance computing and can distinguish important and unimportant details of
the protein structures from the point of view of transport properties.
\end{abstract}

\keywords{Landauer fromula \and conductance of biomolecules \and metallic contacts}

\section{Introduction}

Electron transport measurements via metallic contacts attached to proteins show anomalous properties relative to electron transfer in homologous structures\cite{bostick2018protein,amdursky2014electronic}. 
Borrowing the concept of {\em stylized facts} from economics\cite{kaldor1961capital}, we can introduce here
three simplified presentations of empirical findings:
\begin{itemize}
    \item Conductance measured between metallic electrodes attached well to large protein structure is unexpectedly high. It falls into the nano Siemens scale even over distances of several nano meters\cite{artes2011transistor,zhang2017observation,zhang2019role}.
    \item The conductance does not show significant decay by increasing the distance of the electrodes\cite{yoo2011fabrication,korpany2012conductance,zhang2019electronic}.
    \item The conductance remains nearly constant when temperature is changed from tens of Kelvins to ambient temperatures\cite{sepunaru2011solid}.
\end{itemize}

Bioelectronic measurements with metallic contacts chemically bound to molecules can be regarded as molecular junctions and the
Landauer-Büttiker (LB) formula is one of the best theoretical tools to describe quantum conductance at zero temperature in such systems\cite{lambert2015basic}. 
It expresses the conductance 
in terms of the scattering matrix elements between metallic leads. In the simplest case, only a single scattering channel is open in a narrow lead and a single transmission $T(E_F)$ at the Fermi energy $E_F$ determines the conductance
\begin{equation}
    G=\frac{2e^2}{h}T(E_F),
\end{equation}
where the unit of the quantum conductance $2e^2/h \approx 77481nS$. 
At high temperatures this formula is not applicable and electron transfer is usually
treated in semiclassical Marcus theory\cite{marcus1956theory} (MT) 
\begin{equation}
k_{et} = \frac{2\pi}{\hbar}|H_{AB}|^2 \frac{1}{\sqrt{4\pi \lambda k_BT}}\exp \left ( -\frac{(\lambda +\Delta G^\circ)^2}{4\lambda k_BT} \right ),\label{MT}
\end{equation}
where the electron transfer rate $k_{et}$ is expressed in terms of the electronic coupling between the initial and final states $|H_{AB}|$, the reorganization energy $\lambda$, and the total Gibbs free energy change $\Delta G^\circ$.
Then Nitzan et al. showed\cite{segal2000electron,nitzan2001electron,nitzan2001relationship} that the conductance of molecular junctions is proportional with the electron transfer rate within the same approximation. Since electronic states in biomolecules are highly localized, overlap between distant electronic states decay fast and
both LB and MT yields exponentially decaying conductance $G\sim \exp(-\beta l)$,
where $l$ is the distance of the electrodes and $1/\beta$ is about  $1 \mbox{\normalfont\AA}$. Temperature dependence is also exponential due to the Arrhenius factor in (\ref{MT}).  Both LB and MT are limiting cases only and
the electron-vibrational (electron-phonon) interactions should be treated
more carefully in the intermediate regime. Recently, in Ref.\cite{sowa2018beyond}
this derivation has been carried out for a molecular junction modelled as a single electronic level coupled with a collection of normalized vibrational modes. Using a generalized quantum master equation, it has been shown that LB and MT can be viewed as two limiting cases of this more general expression.

The current impasse in interpreting experimental results is coming from the fact
that charge transport through molecular junctions is described either as a purely coherent or a purely classical phenomenon. 
In recent years it became clear that decoherence plays an important role in biological energy transfer processes \cite{sarovar2010quantum,engel2007evidence} and these effects are not covered by the semiclassical approximation\cite{mohseni2008environment}. In this paper we show that decoherence due to strong coupling to vibrational modes plays an
important role in electron transport processes as well.
We generalize the LB formula for conditions relevant in bioelectronic systems operating at strong decoherence. We capture new physics, which is absent in both limiting cases but plays a significant role when a metallic electrode is attached to the molecule and the chemical bonding is strong between the metal and the nearest localized electronic state of the molecule while direct tunneling between the two distant localized electronic states is exponentially suppressed. 

Our starting point is the derivation of low temperature LB formula for molecules by Datta et al. in Refs.\cite{zahid2003electrical,datta1997electronic} which we summarize here briefly. The molecule is coupled to a left and a right electrode.
The discrete levels of the molecule $\varepsilon_n$ are non-resonantly coupled to to the left
and right electrodes with coupling strengths $\Gamma_n^L$ and $\Gamma_n^R$ respectively. The presence of contacts
broadens the levels and can be described with a Lorentzian density of states
\begin{equation}
d_n(E)=\frac{1}{2\pi}\frac{\Gamma_n}{(E-\varepsilon_n)^2+\Gamma^2_n/4},
\end{equation}
where $\Gamma_n=\Gamma_n^L+\Gamma_n^R$ is the broadening due to the contacts. If the level $\varepsilon_n$ were in equilibrium with the left contact then the number of electrons
$N^L_n$, occupying the level would be given by
\begin{equation}
N^{L}_n=2\int_{-\infty}^{+\infty}d_n(E)f(E,\mu_{L})dE,
\end{equation}
where $\mu_{L}$ is the chemical potential in the left lead, $f_e(E,\mu)=(1+e^{(E-\mu)/kT})^{-1}$ is the Fermi distribution and the factor 2 stands for
spin degeneracy. A similar expression is valid for $N^R_n$ when the molecule is in equilibrium with the right lead. Under non-equilibrium conditions the number of
electrons $N_n$ will be somewhere in between $N_n^L$ and $N_n^R$ and we can write the net
current at the left junction as
\begin{equation}
I_n^L=\frac{e\Gamma_n^L}{\hbar}(N_n^L-N_n),
\end{equation}
where $\Gamma_n^L/\hbar$ is the escape rate from the level to the left lead.
Similarly, for the right junction
\begin{equation}
I_n^R=\frac{e\Gamma_n^R}{\hbar}(N_n-N_n^R).
\end{equation}
Steady state requires $I_n^L=I_n^R$ yielding $
N_n=(\Gamma_n^LN_n^L+\Gamma_n^RN_n^R)/(\Gamma_n^L+\Gamma_n^R)$. The current through the level is then
\begin{equation}
I_n=I_n^L=I_n^R=\frac{e}{\hbar}\frac{\Gamma_n^L\Gamma_n^R}{\Gamma_n^L+\Gamma_n^R}(N_n^L-N_n^R).
\end{equation}
Using the density of states the current trough the molecule can be written as
\begin{equation}
I=\sum_n I_n=\sum_n\frac{2e}{h}\int_{-\infty}^{+\infty}\frac{\Gamma_n^L\Gamma_n^R}{(E-\varepsilon_n)^2+(\Gamma_n^L+\Gamma_n^R)^2/4}(f(E,\mu_{L})-f(E,\mu_{R}))dE.
\end{equation}
In the linear regime and at low temperatures $kT\rightarrow 0$ the chemical potential in the left and right electrodes  is $\mu_{L/R}=E_F\pm eU/2$, the difference  $ f(E,\mu_{L})-f(E,\mu_{R})\approx \delta(E-E_F)eU$ and then
\begin{equation}
I=\frac{2e^2}{h}T(E_F)\cdot U,
\end{equation}
where the transmission is given by the Breit-Wigner formula\cite{lambert2015basic}
\begin{equation}
T(E_F)=\sum_n\frac{\Gamma_n^L\Gamma_n^R}{(E_F-\varepsilon_n)^2+(\Gamma_n^L+\Gamma_n^R)^2/4}.
\end{equation}
The formula is valid when the Fermi energy $E_F$ is close to an eigenenergy of the isolated molecule and  the level spacing of the isolated molecule is larger than $\Gamma_n^L+\Gamma_n^R$.
When energy of the isolated molecule $\varepsilon_n$ is above the Fermi energy $E_F$
the expression 
\begin{equation}
\frac{\Gamma_n^L\Gamma_n^R}{(E_F-\varepsilon_n)^2+(\Gamma_n^L+\Gamma_n^R)^2/4}.
\end{equation}
describes electron transmission. An electron tunneling trough an unoccupied orbital of the molecule from the left
lead to the right one when electric field is switched on in that direction. When
$\varepsilon_n$ is below the Fermi energy $E_F$ the orbital is occupied and  $E_F-\varepsilon_n$ is positive.
This case describes hole transport. A positively charged hole tunneling trough the molecule from the right to the left electrode with negative tunneling energy $\varepsilon_n-E_F$. This way, both processes contribute to the net current with the same sign.  In the next section we generalize Datta's result for finite temperatures.

\section{Derivation of a Landauer formula for bioelectronics}

The derivation of the LB formula in the Breit-Wigner approximation is an
especially suitable starting point for generalization to include vibronic effects.
When such effects are present, the electron (or hole) which tunnels into an orbital of the molecule
is able to transit to another orbital of the molecule since the energy difference between the orbitals
can be taken away (or supplied) by the interaction with the vibrational modes. Note that even at zero temperature
the electron can hop to lower energy so vibronic effects modify the LB formula even in that case.
The steady state condition $I_n^L=I_n^R$ should be modified to account for hoping in and
out of an orbital. Electrons can hop between the (nearly unoccupied) electronic states above the Fermi energy while holes can hop between the (nearly occupied) states below it. Accordingly, we should treat electrons and holes separately. For brevity, we derive the results for electrons in detail and then give the analogous expressions for holes.

Quantum master equations are the most convenient way to describe the transition between electronic states.
They are in general non-Markovian, but for practical purposes can be approximated with Markovian equations 
such as the Redfield equation\cite{breuer2002theory}.
The reduced density matrix elements of the molecule in the energy basis $\varrho_{nm}$ then satisfy a linear equation
\begin{equation}
    \partial_t \varrho_{nm}=\frac{i}{\hbar}(\varepsilon_m-\varepsilon_n+i\Gamma_n/2+i\Gamma_m/2)\varrho_{nm}+\sum_{kl}R_{nmkl}\varrho_{kl}+J_n\delta_{nm},\label{redfield}
\end{equation}
where $J_n$ is the external current,$\varepsilon_n+i\Gamma_n/2$ is the broadened level and 
$R_{nmkl}$ is the Bloch-Redfield tensor describing transitions due to the couplings to the phononic vibrations.
The tetradic matrix $R_{mnkl}$ is the transfer rate from $\varrho_{kl}$ to $\varrho_{mn}$
and can be expressed as
\begin{equation}
    R_{mnkl}=\Gamma_{lmnk}+\Gamma_{knml}^*-\delta_{ml}\sum_p\Gamma_{nppk}-\delta_{nk}\sum_q\Gamma_{mqql}^*,\label{R}
\end{equation}
where 
$$
\Gamma_{mnkl}=\frac{1}{\hbar^2}\int_0^\infty d\tau e^{-i(\varepsilon_k-\varepsilon_l)\tau/\hbar}\langle \boldsymbol{V}_{nm}(\tau) \boldsymbol{V}_{kl}(0) \rangle_b,
$$
are Fourier-Laplace transforms of correlation functions of
matrix elements $\boldsymbol{V}_{ij}$ of the system-bath coupling operator between system eigenstates $i$ and $j$
and the brackets represent a trace over the thermalized bath. We note that in this level of description the electron and
hole states do not mix. Electrons can hop on states above, while holes below the Fermi energy. Consequently, there
are two separate Redfield equations, one for the electrons and one for the holes. The four indices of $R_{nmkl}$ and the
two indices of $\varrho_{nm}$ should be either all electron or hole states.  

We normalize this equation such a way that the diagonal elements of the density matrix can correspond to the 
occupations $\varrho_{nn}=N_n$ introduced in the previous section. In this case the external (material) current
becomes
\begin{equation}
    J_n=\frac{\Gamma_n^L}{\hbar}\varrho_n^L+\frac{\Gamma_n^R}{\hbar}\varrho_n^R,
    \label{current}
\end{equation}
where 
\begin{equation}
\varrho^{L/R}_n=2\int_{-\infty}^{+\infty}d_n(E)f(E,\mu_{L/R})dE,
\end{equation}
is the occupation of the levels when the molecule is in equilibrium with the left/right lead.   

In absence of an electric field ($U=0$) the system is in equilibrium and $\mu_{L/R}=E_F$.
The Fermi energy is between the HOMO and the LUMO  energies
$\varepsilon_{N/2}=\varepsilon_{HOMO} <E_F < \varepsilon_{N/2+1}=\varepsilon_{LUMO}$.
The number of electrons $N$ in the molecule is given by the sum of occupancies
$N=\sum_{n=1}^{\infty} 2f_e(\varepsilon_n,E_F)$. This can be written also as
\begin{equation}
\sum_{n=1}^{N/2} (1-f(\varepsilon_n,E_F))=\sum_{n=N/2+1}^{\infty}f_e(\varepsilon_n,E_F),
\end{equation}
and with the Fermi distribution for holes $f_h(E,\mu)=1-f_e(E,\mu)=(1+e^{(\mu-E)/kT})^{-1}$
it can be simplified to
\begin{equation}
\sum_{n=1}^{N/2} f_h(\varepsilon_n,E_F)=\sum_{n=N/2+1}^{\infty}f_e(\varepsilon_n,E_F),\label{sumrule}
\end{equation}
meaning that the number of holes below the Fermi energy is the same as the electrons
above it and the molecule is charge neutral.
An important aspect of bioelectronic systems is that they have highly localized electronic states and a large HOMO-LUMO gap ($\sim eV$) in accordance with standard Density Functional Theory (DFT) calculations\cite{lever2013electrostatic}.
Based on this we can assume that $e^{(E_F-\varepsilon_n)/kT}\ll 1$ for electronic states and also
$e^{(\varepsilon_n-E_F)/kT}\ll 1$ for hole states and we can replace the Fermi distribution with the
Boltzmann distribution in both cases $f_e(E,\mu)=e^{-(E-\mu)/kT}$ and $f_h(E,\mu)=e^{-(\mu-E)/kT}$.
Accordingly, the equilibrium occupancy for electrons is given by the Boltzmann distribution
\begin{equation}
    \overline{\varrho}^{L/R}_n= 
    \begin{cases} 
    2e^{-(\varepsilon_n-E_F)/kT} \: \mbox{for electrons},\\ 
    2e^{-(E_F-\varepsilon_n)/kT} \: \mbox{for holes}.
    \end{cases}
\end{equation}
Using (\ref{sumrule}) in the Boltzmann approximation we can introduce
the partition function 
$$Z(T)=\sum_{n=1}^{N/2} e^{-(E_F-\varepsilon_n)/kT} = \sum_{n=N/2+1}^{\infty}e^{-(\varepsilon_n-E_F)/kT},$$
which is the same for electrons and holes.

In absence of electric field the steady state solution of the Redfield equation (\ref{redfield})
is also the Boltzmann distribution 
\begin{equation}
    \overline{\varrho}_{nn}= 
    \begin{cases} 
    2e^{-(\varepsilon_n-E_F)/kT} \: \mbox{for electrons},\\ 
    2e^{-(E_F-\varepsilon_n)/kT} \: \mbox{for holes}.
    \end{cases}
\end{equation}
Then in equilibrium each term vanishes separately in
\begin{equation}
I=e\sum_n\frac{\Gamma_n^L}{\hbar}(\varrho^L_n-\overline{\varrho}_{nn})=0,
\end{equation}
and there is no electric current.

When the electric field is switched on ($U\neq 0$) the system is out of equilibrium. In the linear
regime we can expand the deviation from the equilibrium 
\begin{equation}
\varrho^{L/R}_n=2\int_{-\infty}^{+\infty}dEd_n(E)[f(E,E_F)\mp f'(E,E_F)eU/2 + ...]\approx \overline{\varrho}^{L/R}_n \pm D_n(E_F,T)eU,
\end{equation}
where
\begin{equation}
D_n(E_F,T)=-\int_{-\infty}^{+\infty} f'(E,E_F)d_n(E)dE.
\end{equation}
We can introduce the deviation of the density matrix elements from
their equilibrium value $\varrho_{nm}'=\varrho_{nm}-\overline{\varrho}_{nm}$ and using (\ref{redfield}) and (\ref{current}) we can write the steady state
equation 
\begin{equation}
-\frac{\Gamma_n^L-\Gamma_n^R}{\hbar}D_n(E_F,T)eU\delta_{nm}=\frac{i}{\hbar}(\varepsilon_m-\varepsilon_n+i\Gamma_n/2+i\Gamma_m/2)\varrho_{nm}'+\sum_{kl}R_{nmkl}\varrho_{kl}',\label{eq20}
\end{equation}
where we grouped the external current to the left hand side.
The current through the molecule is
\begin{equation}
I=e\sum_n\frac{\Gamma_n^L}{\hbar}(D_n(E_F,T)eU-\varrho_{nn}').
\label{eq21}
\end{equation}
To get the general solution of (\ref{eq20}) we can introduce the tetradic matrix $L_{nmkl}=(i/\hbar)(\varepsilon_m-\varepsilon_n+i\Gamma_n/2+i\Gamma_m/2)\delta_{nk}\delta_{ml}+R_{nmkl}$ and can write
\begin{equation}
-\frac{\Gamma_n^L-\Gamma_n^R}{\hbar}D_n(E_F,T)eU\delta_{nm}=\sum_{kl}L_{nmkl}\varrho_{kl}'.
\end{equation}
The solution of this equation can be given in terms of the inverse matrix
\begin{equation}
\varrho_{nm}'=-\sum_kL^{-1}_{nmkk}\frac{\Gamma_k^L-\Gamma_k^R}{\hbar}D_k(E_F,T)eU,\label{solution}
\end{equation}
where the inverse satisfies the relation $\sum_{pq}L_{nmpq}L^{-1}_{pqkl}=\delta_{nk}\delta_{ml}$. Substituting this solution into (\ref{eq21}) we get the generalized Landauer-Büttiker formula  
\begin{equation}
I=\frac{e^2U}{\hbar}\sum_n D_n(E_F,T) \left[\Gamma_n^L+\frac{1}{\hbar}\sum_k \Gamma_k^L L^{-1}_{kknn}(\Gamma_n^L-\Gamma_n^R)\right].\label{genLB}
\end{equation}
This formula can be brought (see Appendix A) to a form which reflects the left-right symmetry 
\begin{equation}
G=\frac{e^2}{2\hbar}\sum_n D_n(E_F,T)\left[\Gamma_n^L+\Gamma_n^R+\frac{1}{\hbar}\sum_k (\Gamma_k^L-\Gamma_k^R) L^{-1}_{kknn}(\Gamma_n^L-\Gamma_n^R)\right],\label{conductance}
\end{equation}
which is our main result. We note that $L^{-1}_{kknn}=0$ unless both $k$ and $n$ are electron or hole states, consequently the conductance can be split to an electron and a hole part $G=G_e+G_h$, where
\begin{equation}
    G_e=\frac{e^2}{2\hbar}\sum_{n=N/2+1}^{\infty} D_n(E_F,T)\left[\Gamma_n^L+\Gamma_n^R+\frac{1}{\hbar}\sum_{k=N/2+1}^{\infty} (\Gamma_k^L-\Gamma_k^R) L^{-1}_{kknn}(\Gamma_n^L-\Gamma_n^R)\right],\label{econd}
\end{equation}
and
\begin{equation}
    G_h=\frac{e^2}{2\hbar}\sum_{n=1}^{N/2} D_n(E_F,T)\left[\Gamma_n^L+\Gamma_n^R+\frac{1}{\hbar}\sum_{k=1}^{N/2} (\Gamma_k^L-\Gamma_k^R) L^{-1}_{kknn}(\Gamma_n^L-\Gamma_n^R)\right],\label{hcond}
\end{equation}
just like in the zero temperature LB formula discussed before.

\section{Electron transfer}

The present formalism allows us to calculate the electron transfer along the same lines. For specificity, the left electrode plays the role of donor and the right 
electrode the acceptor site. The electron charge on the donor and acceptor sites
follows the Fermi distribution, which can be approximated by the Boltzmann distribution due to the large HOMO-LUMO gap. The electron-hole picture is useful
here as well. Electrons traversing the molecule via almost unoccupied orbitals above the Fermi energy contribute to the electron part, while transfer via almost fully occupied orbitals below the Fermi level can be regarded as hole transport.  
Introducing $\varrho^D$ for the total density on the left electrode and
$\varrho^A$ for the right electrode, the left and right densities become
$\varrho^{L/R}_n=\varrho^{D/A}p_n^B,$ where
\begin{equation}
    p_n^B= 
    \begin{cases} 
    2e^{-(\varepsilon_n-E_F)/kT}/Z \: \mbox{for electrons},\\ 
    2e^{-(E_F-\varepsilon_n)/kT}/Z \: \mbox{for holes}.
    \end{cases}
\end{equation}
Then, given the external current 
\begin{equation}
    J_n=\frac{\Gamma_n^L}{\hbar}\varrho_n^L+\frac{\Gamma_n^R}{\hbar}\varrho_n^R,
\end{equation}
we have to solve the Redfield equation
\begin{equation}
-J_n\delta_{nm}=\sum_{kl}L_{nmkl}\varrho_{kl},
\end{equation}
to get the total material current
\begin{equation}
J=\sum_n\frac{\Gamma_n^L}{\hbar}(\varrho_n^L-\varrho_{nn}),
\end{equation}
which leads to
\begin{equation}
    J=\frac{1}{\hbar }\varrho_D\sum_n p_n^B\Gamma_n^L \left[1+\frac{1}{\hbar}\sum_k \Gamma_k^L L_{kknn}^{-1}\right]  +\frac{1}{\hbar }\varrho_A\sum_n p_n^B\Gamma_n^L\left[\frac{1}{\hbar}\sum_k \Gamma_k^RL_{kknn}^{-1}\right].
\end{equation}
Electrons can escape from the acceptor site with escape rate $\kappa=J/\varrho^A$ and the electron transfer rate is the ratio of the material current and the density at the donor site $k_{ET}=J/\varrho^D$. We can express the transfer rate with the escape rate
\begin{equation}
    k_{ET}=\frac{\sum_n p_n^B\Gamma_n^L \left[1+\frac{1}{\hbar}\sum_k \Gamma_k^L L_{kknn}^{-1}\right]/\hbar}{  1+(1/\kappa)\sum_n p_n^B\Gamma_n^L\left[1+\frac{1}{\hbar}\sum_k \Gamma_k^LL_{kknn}^{-1}\right]/\hbar},\label{ET}
\end{equation}
where we used the result of Appendix A to show that
\begin{equation}
    \sum_n p_n^B\Gamma_n^L\left[\frac{1}{\hbar}\sum_k \Gamma_k^RL_{kknn}^{-1}\right]=-\sum_n p_n^B\Gamma_n^L \left[1+\frac{1}{\hbar}\sum_k \Gamma_k^L L_{kknn}^{-1}\right].
\end{equation}

\section{Weak contacts}

The strength of the contacts relative to the thermal energy plays a crucial role in the conductance properties of these systems.
When the contacts are weak the electrons and holes can enter the molecule from the lead via thermal excitation. In this case 
the conductance is intimately related to electron transfer as it has been shown in Refs.\cite{nitzan2001electron,nitzan2002relationship,nitzan2001relationship}. Here, we derive an exact formula between electron transfer and electron conductance. 

When the contacts are weak compared to the thermal energy $\Gamma_n\ll kT$ the density of states consists of delta peaks $d_n(E)\approx\delta(E-\varepsilon_n)$, the Fermi distribution is well approximated with the Boltzmann and we get 
\begin{equation}
D_n(E_F,T)\approx-\int_{-\infty}^{+\infty} f'(E,E_F)\delta(E-\varepsilon_n)dE=-f'(\varepsilon_n,E_F)\approx
\begin{cases} 
    e^{-(\varepsilon_n-E_F)/kT}/kT \: \mbox{for electrons},\\ 
    e^{-(E_F-\varepsilon_n)/kT}/kT \: \mbox{for holes}.
    \end{cases}
\end{equation}
This can be written in the more compact form using the normalized Boltzmann distribution $D_n(E_F,T)=(Z(T)/2kT)p_n^B$.
Substituting this into (\ref{genLB}) and using the sum rule derived in Appendix D we can eliminate
$\Gamma_n$ and get the form
\begin{equation}
G=\frac{e^2Z(T)}{kT\hbar}\sum_n p_n^B\Gamma_n^L \left[1+\frac{1}{\hbar}\sum_k \Gamma_k^L L^{-1}_{kknn}\right].
\end{equation}
The sum in this expression appears in (\ref{ET}) as well so that the electron transfer rate can be expressed with the 
conductance directly
\begin{equation}
    k_{ET}=\frac{(kT/e^2 Z(T))G}{  1+(kT/e^2 Z(T))G/\kappa }.
\end{equation}
When the escape from the acceptor is strong, we can neglect $1/\kappa$ and the conductance is proportional with the 
electron transfer rate
\begin{equation}
    G=\frac{e^2}{kT}Z(T)\cdot k_{ET}.
\end{equation}
In biomolecules where the HOMO-LUMO gap $\Delta_{HL}=E_{LUMO}-E_{HOMO}$ is large compared to the thermal energy $kT$ the partition sum is dominated by the
gap $ Z(T)\approx e^{-(E_{LUMO}-E_F)/kT}\approx e^{-\Delta_{HL}/2kT}$ and the conductance is
\begin{equation}
    G=\frac{e^2}{kT} e^{-\Delta_{HL}/2kT} k_{ET},
\end{equation}
where $e^{-\Delta_{HL}/2kT}\sim 10^{-40}$ making these systems practically insulators. In case of bridged molecular systems considered by Nitzan in Ref.\cite{nitzan2002relationship} the partition function is $Z\approx e^{-\Delta E/kT}$, where
$\Delta E =E_B-E_F$ is the difference between the Fermi energy of the metallic leads and the average energy of the bridged
system and we recover Nitzan's formula
\begin{equation}
    G=\frac{e^2}{kT} e^{-\Delta E/kT} k_{ET}.
\end{equation}
Finally, in case of a coherent one dimensional bridged molecule system where the gap is zero $\Delta E=0$ considered in Ref.\cite{segal2000electron} we can use the classical partition function of a free particle in a one dimensional box:
\begin{equation}
    Z=\int_0^{L_M} \int_{-\infty}^{+\infty}\frac{dxdp}{h} e^{-p^2/2mkT}=\frac{L_M\sqrt{2\pi mkT}}{h},
\end{equation}
where $L_M$ is the length of the molecule and $m$ is the effective mass of the electron in the molecule.
In this case we get the formula
\begin{equation}
    G=\frac{2e^2}{h}L_M\sqrt{\frac{\pi m}{2kT}}  k_{ET},
\end{equation}
which is slightly different from the heuristically derived result in Ref.\cite{segal2000electron}
$G=\frac{2e^2}{h}L_M\sqrt{\frac{m}{2E_F}}  k_{ET}$.
The difference comes from the fact that in Ref.\cite{segal2000electron}
it is assumed that the electron traverses the molecule with effective velocity $v_F=\sqrt{2E_F/m}$ and energy
$E_F$, while in reality the electron transfer happens at the thermal energy scale $kT$ and with effective
velocity $v_T=\sqrt{2kT/m}$. The two expressions differ in a factor
of $v_F/v_T\approx 5$ only, which is hard to verify experimentally.

\section{Strong contacts}

In the opposite case, when the contacts are strong, electrons and holes enter the molecule via 
tunneling. This is a new regime not covered by the previous studies and we show that the relation between electron transfer 
rate and the conductance breaks down.

When the thermal energy is small compared to the strengths of the contacts $kT\ll\Gamma_n$ the density of states is smooth and the thermal distribution is approximately $-f'(E,E_F)\approx \delta(E-E_F)$ so that  
\begin{equation}
D_n(E_F,T)=\int_{-\infty}^{+\infty} d_n(E)\delta(E-E_F)dE= \frac{1}{2\pi}\frac{\Gamma_n}{(E_F-\varepsilon_n)^2+\Gamma^2_n/4}.
\end{equation}
Substituting this into (\ref{econd}) and (\ref{hcond}) we get the electron
\begin{equation}
    G_e=\frac{e^2}{2h}\sum_{n=N/2+1}^{\infty}\frac{\Gamma_n}{(E_F-\varepsilon_n)^2+\Gamma^2_n/4} \left[\Gamma_n^L+\Gamma_n^R+\frac{1}{\hbar}\sum_{k=N/2+1}^{\infty} (\Gamma_k^L-\Gamma_k^R) L^{-1}_{kknn}(\Gamma_n^L-\Gamma_n^R)\right], \label{sronge}
\end{equation}
and the hole conductance
\begin{equation}
    G_h=\frac{e^2}{2h}\sum_{n=1}^{N/2} \frac{\Gamma_n}{(E_F-\varepsilon_n)^2+\Gamma^2_n/4}\left[\Gamma_n^L+\Gamma_n^R+\frac{1}{\hbar}\sum_{k=1}^{N/2} (\Gamma_k^L-\Gamma_k^R) L^{-1}_{kknn}(\Gamma_n^L-\Gamma_n^R)\right].\label{strongh}
\end{equation}
In this case, tunneling populates the levels and the relation with the Boltzmann distribution
breaks down. It is no longer possible to connect electron transfer and conductance with a simple formula.

The other equally crucial factor in the conductance of these systems is the strength of the coupling to the environment through the vibrational degrees of freedom, which is encoded in the matrix $L_{kknn}^{-1}$. In general it is complicated to calculate this quantity since the microscopic details of the couplings between the electron and vibration degrees of freedom can play an
important role. To get an insight we consider the two limiting cases, when the coupling to the vibrations is negligible (coherent case) and the opposite case, when coupling to the environment dominates (full decoherence). Surprisingly, in the latter case the details of the coupling drop out and the conductance depends on the contact strengths and the energy spectrum
of the molecule as we show next.

In the coherent case, when the Redfield tensor elements describing the coupling to the heath bath are small ($|R|\ll \Gamma$)  we can neglect them and the inverse operator matrix elements become
\begin{equation}
    L_{nnmm}^{-1}=-\frac{\hbar\delta_{nm}}{\Gamma_n^L+\Gamma^R_n}.\label{lnm}
\end{equation}
Substituting this into (\ref{econd}) and (\ref{hcond}) we get the electron
\begin{equation}
    G_e=\frac{2e^2}{h}\sum_{n=N/2+1}^{\infty}D_n(E_F,T) \frac{\Gamma_n^L\Gamma_n^R}{\Gamma_n^L+\Gamma_n^R},
\end{equation}
and the hole conductance
\begin{equation}
    G_h=\frac{2e^2}{h}\sum_{n=1}^{N/2} D_n(E_F,T)\frac{\Gamma_n^L\Gamma_n^R}{\Gamma_n^L+\Gamma_n^R}.
\end{equation}
In case of strong contacts and ($|R|, kT \ll \Gamma$) we can substitute this 
and recover the Landauer-Büttiker formula in the Breit-Wigner approximation
\begin{equation}
G=G_e+G_h=\frac{2e^2}{h}\sum_n\frac{\Gamma_n^L\Gamma_n^R}{(E_F-\varepsilon_n)^2+(\Gamma_n^L+\Gamma_n^R)^2/4}.
\end{equation}

\subsection{Strong and weak contact mixed}

It is an important experimental situation\cite{zhang2017observation,zhang2019role}, when one of the contacts is strong and the other one is weak. In some cases,
the strongly coupled electrode forms a covalent bond with a specific atom of the molecule, while the other electrode is coupled
non-specifically with weak coupling. Interestingly, to meet the condition of the strong contact case $kT\ll\Gamma_n=\Gamma_n^L+\Gamma_n^R$ it is sufficient if only one of the contacts is strong. For example, if the left contact is strong $kT\ll\Gamma_n^L$ and the right contact is weak we can neglect $\Gamma_n^R$ in (\ref{sronge}) and (\ref{strongh}) 
and arrive at the expression for the electron
\begin{equation}
    G_e=\frac{e^2}{2h}\sum_{n=N/2+1}^{\infty}\frac{\Gamma_n^L}{(E_F-\varepsilon_n)^2+(\Gamma^L_n)^2/4} \left[\Gamma_n^L+\frac{1}{\hbar}\sum_{k=N/2+1}^{\infty} \Gamma_k^L L^{-1}_{kknn}\Gamma_n^L\right], \label{srongweake}
\end{equation}
and for the hole conductance
\begin{equation}
    G_h=\frac{e^2}{2h}\sum_{n=1}^{N/2} \frac{\Gamma_n^L}{(E_F-\varepsilon_n)^2+(\Gamma^L_n)^2/4}\left[\Gamma_n^L+\frac{1}{\hbar}\sum_{k=1}^{N/2} \Gamma_k^L L^{-1}_{kknn}\Gamma_n^L\right].\label{strongweakh}
\end{equation}
This means that high conductance can arise not just between two strong contacts but also in the single strong contact case.
We discuss this possibility further in the next sections.

\subsection{Strong decoherence}

When the Bloch-Redfield terms are small compared to the couplings to the leads we can expect just some moderate deviations from the LB formula. Interesting new physics arises in the opposite case  ($\Gamma \ll |R|$), when the electrons or holes arriving from the leads are strongly mixed in the molecule. In absence of the coupling to the leads the matrix
\begin{equation}
L_{nmkl}^0=(i/\hbar)(\varepsilon_m-\varepsilon_n)\delta_{nk}\delta_{ml}+R_{nmkl},\label{L0}
\end{equation}
describes the isolated molecule. The steady state solution of the density matrix of this system is the Boltzmann distribution. The normalized equilibrium density matrix is
$\varrho_{nn}^0=e^{-(\varepsilon_n-E_F)/kT}/Z(T)$ for
electrons and $\varrho_{nn}^0=e^{-(E_F-\varepsilon_n)/kT}/Z(T)$ for holes. The inverse operator can be calculated perturbatively
in the limit of small coupling $\Gamma_n \rightarrow 0$. For the details of the calculation see Appendix B and the diagonal elements
of (\ref{a14}) become
\begin{equation}
    L_{nnmm}^{-1}\approx -\frac{\hbar\varrho_{nn}^0}{\sum_p \Gamma_p\varrho_{pp}^0},\label{a14}
\end{equation}
where all the indices should be either electrons or holes.
Substituting this into (\ref{conductance}) yields the hole conductance
\begin{equation}
G_h=\frac{e^2}{h}\sum_{n=1}^{N/2} \frac{\Gamma_n}{(E_F-\varepsilon_n)^2+\Gamma^2_n/4}  \frac{\Gamma_n^R \langle\Gamma^L_h(T)\rangle +\Gamma_n^L\langle\Gamma^R_h(T)\rangle}{\langle\Gamma^L_h(T)\rangle+\langle\Gamma^R_h(T)\rangle},\label{Gvibrh}
\end{equation}
and the electron conductance
\begin{equation}
G_e=\frac{e^2}{h}\sum_{N/2+1}^{\infty} \frac{\Gamma_n}{(E_F-\varepsilon_n)^2+\Gamma^2_n/4}  \frac{\Gamma_n^R \langle\Gamma^L_e(T)\rangle +\Gamma_n^L\langle\Gamma^R_e(T)\rangle}{\langle\Gamma^L_e(T)\rangle+\langle\Gamma^R_e(T)\rangle},\label{Gvibre}
\end{equation}
such that the total conductance is the sum of the two
\begin{equation}
    G=G_h+G_e.
\end{equation}
We introduced the weighted sums of the left and the right coupling strengths for electron 
\begin{equation}
    \langle\Gamma^{L/R}_e(T)\rangle=\sum_{n=N/2+1}^{\infty} \Gamma_n^{L/R}e^{-(\varepsilon_n-E_F)/kT},\label{WS}
\end{equation} 
and for the hole states
\begin{equation}
    \langle\Gamma^{L/R}_h(T)\rangle=\sum_{n=1}^{N/2} \Gamma_n^{L/R}e^{-(E_F-\varepsilon_n)/kT}.
\end{equation} 
We can carry out the summation and get the conductance in the vibration dominated regime
\begin{equation}
G=\frac{e^2}{h}\left[\frac{ T^R_h(E_F) \langle\Gamma^L_h(T)\rangle + T^L_h(E_F)\langle\Gamma^R_h(T)\rangle}{\langle\Gamma^L_h(T)\rangle+\langle\Gamma^R_h(T)\rangle}+\frac{ T^R_e(E_F) \langle\Gamma^L_e(T)\rangle + T^L_e(E_F)\langle\Gamma^R_e(T)\rangle}{\langle\Gamma^L_e(T)\rangle+\langle\Gamma^R_e(T)\rangle}\right],\label{T}
\end{equation}
where we introduced the sums for holes
\begin{equation}
  T^{L/R}_h(E_F)  = \sum_{n=1}^{N/2} \frac{\Gamma_n \Gamma_n^{L/R}}{(E_F-\varepsilon_n)^2+\Gamma^2_n/4},
\end{equation}
and for electrons
\begin{equation}
  T^{L/R}_e(E_F)  = \sum_{n=N/2+1}^{\infty} \frac{\Gamma_n \Gamma_n^{L/R}}{(E_F-\varepsilon_n)^2+\Gamma^2_n/4}.
\end{equation}
A remarkable property of this new conductance formula is that it is independent of the details of the vibrational process and relies solely on the equilibrium distribution and the couplings to the leads. 
The ratio 
\begin{equation}
    P^L_h(T)=\frac{\langle\Gamma^{L}_h(T)\rangle}{\langle\Gamma^{L}_h(T)\rangle+\langle\Gamma^{R}_h(T)\rangle},\label{P}
\end{equation}
in the expression of the conductance can be interpreted as the probability that a hole entering anywhere into the molecule leaves
it towards the left lead. The sum $T_h^R(E_F)$ is the probability that a hole tunnels into the molecule from the right lead. The
product $T^R_h(E_F) \langle\Gamma^L_h(\beta)\rangle$ is the probability that a hole entering from the right lead leaves the
molecule trough the left lead. The four terms in the formula
\begin{equation}
G=\frac{e^2}{h}\left[ T^R_h(E_F) P^L_h(T) + T^L_h(E_F)P^R_h(T)+T^R_e(E_F) P^L_e(T) + T^L_e(E_F)P^R_e(T)\right],\label{formula}
\end{equation}
represent the four scenarios in which electrons and holes can generate current. It has a very modest dependence on temperature and on the distance between the contacts as we show in the next sections. 

Finally, here we can also discuss the sub-case when the left electrode forms a strong specific bond with the molecule while the other electrode is coupled weakly or non-specifically. We can then neglect $T^R_e(E_F)$ and $T^R_h(E_F)$ in (\ref{formula}) and get the simplified expression
\begin{equation}
G=\frac{e^2}{h}\left[ T^L_h(E_F)P^R_h(T)+ T^L_e(E_F)P^R_e(T)\right].\label{strongweakformula}
\end{equation}
This means that electrons and holes can tunnel into the molecule via the strong left contact and some of them can leave
trough the weak contact with the equilibrium probabilities $P^R_e(T)$ and $P^R_h(T)$.

\subsection{Temperature dependence}

The temperature dependence of the conductance is coming from the probabilities. In Appendix C we show that they are 
temperature independent if $kT\ll\varepsilon_{HOMO}-\varepsilon_{HOMO-1}$ and $kT\ll\varepsilon_{LUMO+1}-\varepsilon_{LUMO}$,
which is usually holds in proteins, where the level spacings are typically in the order of $0.1-1.0$ eV and the experimental temperatures are
in the $kT=0.00001-0.025$ eV range. In certain cases, due to the fluctuation of level spacing it can happen that  $\varepsilon_{HOMO}-\varepsilon_{HOMO-1}$ or $\varepsilon_{LUMO+1}-\varepsilon_{LUMO}$ is somewhat lower accidentally,
therefore we keep the temperature dependent terms in leading order to account for these effects. Using the probabilities 
derived in Appendix C we get the following expression for the temperature independent part of the conductance 
\begin{equation}
G_0=\frac{e^2}{h}\left[  \frac{T^R_h(E_F)\Gamma_{HOMO}^{L}+T^L_h(E_F)\Gamma_{HOMO}^{R}}{\Gamma_{HOMO}^{L}+\Gamma_{HOMO}^R} +\frac{T^R_e(E_F)\Gamma_{LUMO}^{L}+T^L_e(E_F)\Gamma_{LUMO}^{R}}{\Gamma_{LUMO}^{L}+\Gamma_{LUMO}^R}\right],\label{tempidG}
\end{equation}
and for the temperature dependence in leading order
\begin{eqnarray}
G_T&=&\frac{e^2}{h}
(T^R_h(E_F)-T^L_h(E_F))\frac{\Gamma_{HOMO}^{L}\Gamma_{HOMO}^{R}}{(\Gamma_{HOMO}^{L}+\Gamma_{HOMO}^R)^2}\left[\frac{\Gamma_{HOMO-1}^{L}}{\Gamma_{HOMO}^L}-   \frac{\Gamma_{HOMO-1}^{R}}{\Gamma_{HOMO}^R}     \right]e^{-(\varepsilon_{HOMO}-\varepsilon_{HOMO-1})/kT} \nonumber \\
&+&\frac{e^2}{h}(T^R_e(E_F)-T^L_e(E_F))\frac{\Gamma_{LUMO}^{L}\Gamma_{LUMO}^{R}}{(\Gamma_{LUMO}^{L}+\Gamma_{LUMO}^R)^2}\left[\frac{\Gamma_{LUMO+1}^{L}}{\Gamma_{LUMO}^L}-   \frac{\Gamma_{LUMO+1}^{R}}{\Gamma_{LUMO}^R}    \nonumber \right]e^{-(\varepsilon_{LUMO+1}-\varepsilon_{LUMO})/kT},\label{tempdpG}
\end{eqnarray}
and $G=G_0+G_T$. The sign of the temperature dependent part is determined by the combined effect of the sign of $\Gamma_{HOMO-1}^{L}/\Gamma_{HOMO}^L-\Gamma_{HOMO-1}^{R}/\Gamma_{HOMO}^R$, $\Gamma_{LUMO+1}^{L}/\Gamma_{LUMO}^L-   \Gamma_{LUMO+1}^{R}/\Gamma_{LUMO}^R$, $T^R_h(E_F)-T^L_h(E_F)$ and $T^R_e(E_F)-T^L_e(E_F)$, which depends not just on whether the left or the
right side is coupled stronger, but also from the details of the couplings to the HOMO vs. HOMO-1 and LUMO vs. LUMO-1.

\subsection{Distance dependence}

Looking at the formula (\ref{formula}) we can realize that neither $T$ nor $P$ has a systematic dependence on the distance of the electrodes. The orbitals of large molecules are localized.  Assuming that the electrodes are far from each other, they are coupled the most strongly
to some localized orbital of the molecule near the electrode. These orbitals do not overlap and direct tunneling is negligible. The electrons
are transported due to the strong vibrational effect. Due to the strong mixing inside of the molecule, the electron moves ergodically inside and looses information about its point of arrival. The exit direction (left or right) is determined solely by the escape rates from the 
molecule. The probability that we find the electron on an orbital is given by the Boltzmann distribution $e^{-(E_F-\varepsilon_n)/kT}/Z(T)$
and the rate of exit from this state to the left electrode is the escape rate multiplied with the probability $(\Gamma^L_n/\hbar)e^{-(E_F-\varepsilon_n)/kT}/Z(T)$. The total rate of escape to the left electrode is $\sum_n(\Gamma^L_n/\hbar)e^{-(E_F-\varepsilon_n)/kT}/Z(T)$ and the probability that the electron leaves the molecule trough the left exit is
the ratio of the total rate of exit to the left divided by the total rate of exit (left+right) $\sum_n(\Gamma_n/\hbar)e^{-(E_F-\varepsilon_n)/kT}/Z(T)$. The result is independent of the distance of the electrodes. 
Looking at one of the sums when the electrodes are far away from each other
\begin{equation}
  T^{L}_h(E_F)  = \sum_{n=1}^{N/2} \frac{\Gamma_n \Gamma_n^{L}}{(E_F-\varepsilon_n)^2+\Gamma^2_n/4},
\end{equation}
we can realize that the strongest contributions come from large $\Gamma_n^{L}$. However, in this case the corresponding $\Gamma_n^{R}$
is very small, since the left and the right electrodes can't couple strongly to the same localized state at the same time. So, when we
calculate this sum, we can drop the terms related to the right electrode and get
\begin{equation}
  T^{L}_h(E_F)  \approx \sum_{n=1}^{N/2} \frac{(\Gamma_n^{L})^2}{(E_F-\varepsilon_n)^2+(\Gamma^L_n)^2/4},
\end{equation}
which depends only on the couplings of the left electrode and is independent of the relative position of the two electrodes.
Similar arguments are true for the right electrode, 
and it is distance independent as well. In summary, none of the terms in formula (\ref{formula}) show any systematic distance
dependence.

\subsection{Order of magnitude}

We can make a rough estimate of the conductance in a typical arrangement. The probabilities $P$ are in order of unity and the typical
value of $T$ determines the order of magnitude. The magnitude of $|E_F-\varepsilon_n|$ is bigger than half of the HOMO-LUMO gap and 
can be $5-10$ eV typically. The couplings $\Gamma_n$ are in the $0.1$ eV range. The ratios $(\Gamma_n/|E_F-\varepsilon_n|)^2$ are then
typically $10^{-4}-10^{-6}$ and $e^2/h\approx 38,000 nS$. The resulting conductances then should be typically in the $10.0-0.01 nS$ range.  

\subsection{Electron transfer at strong decoherence}

It is instructive to calculate the temperature dependence of the electron transfer rate in the strong decoherence case. Using the same approximation yields
\begin{equation}
    k_{ET}=\frac{Z(T)}{\hbar} \left[\frac{\langle\Gamma^{L}_e(T)\rangle \langle\Gamma^{R}_e(T)\rangle}{\langle\Gamma^{L}_e(T)\rangle + \langle\Gamma^{R}_e(T)\rangle}+\frac{\langle\Gamma^{L}_h(T)\rangle \langle\Gamma^{R}_h(T)\rangle}{\langle\Gamma^{L}_h(T)\rangle + \langle\Gamma^{R}_h(T)\rangle}\right],
\end{equation}
where we took the large escape rate $\kappa\rightarrow \infty$ limit in (\ref{ET}). Since the HOMO-LUMO gap is much larger than the thermal energy, we can again keep only the leading terms and get
\begin{equation}
    k_{ET}\approx\frac{e^{-\Delta_{HL}/kT}}{\hbar} \left[\frac{\Gamma^{L}_{LUMO} \Gamma^{R}_{LUMO}}{\Gamma^{L}_{LUMO} + \Gamma^{R}_{LUMO}}+\frac{\Gamma^{L}_{HOMO} \Gamma^{R}_{HOMO}}{\Gamma^{L}_{HOMO} + \Gamma^{R}_{HOMO}}\right],\label{ETtemp}
\end{equation}
where $\Delta_{HL}$ is the HOMO-LUMO gap. This expression shows that the electron transfer rate has an Arrehnius type temperature dependence.  It contains the products
$\Gamma^{L}_{LUMO} \Gamma^{R}_{LUMO}$ and $\Gamma^{L}_{HOMO} \Gamma^{R}_{HOMO}$
which describe tunneling trough the entire molecule and decay exponentially with the size of the molecule. It is obvious that the electron transfer rate is not proportional with the conductance, which - unlike the transfer rate - shows no exponential dependence on distance or temperature.

\section{Experiments}

In this section we show how the present theory explains some of the key experimental findings of protein conductance in the presence of strongly coupled electrodes.

\subsection{Temperature dependence}

\begin{figure}[htb]
\centering
\includegraphics[width=13 cm]{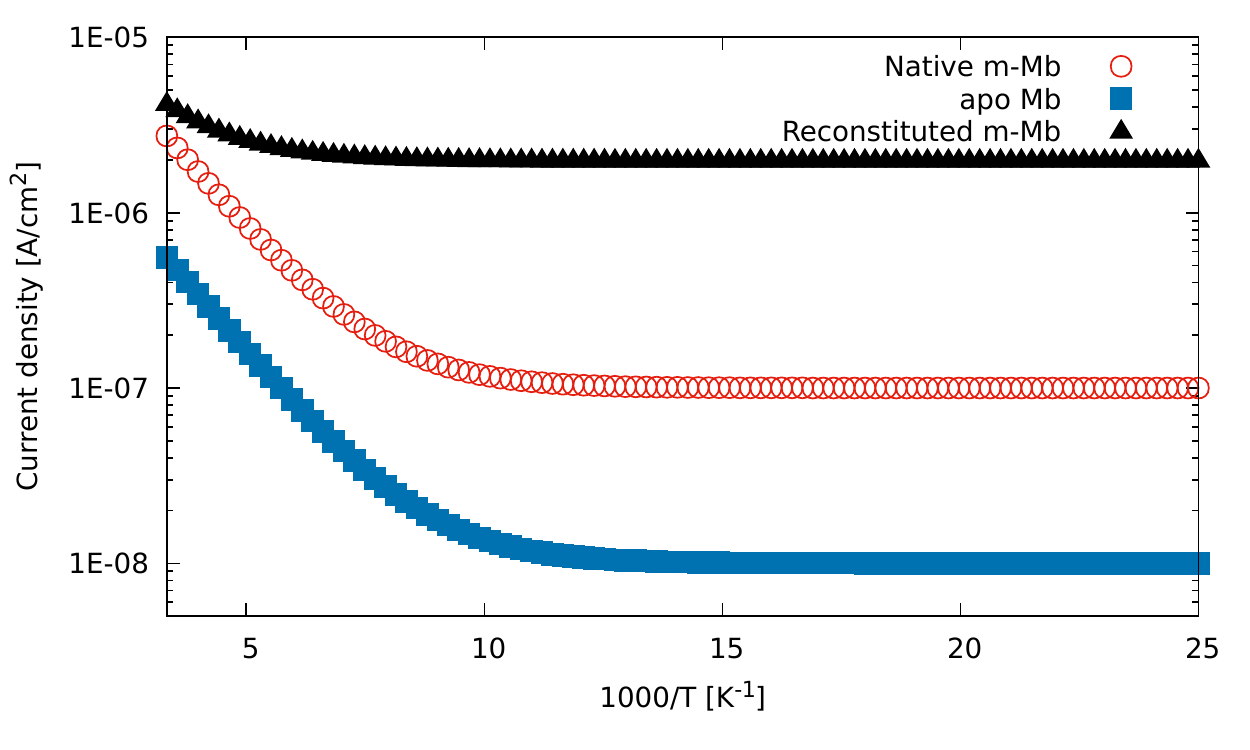}
\caption{Current density curves of experiment Ref\cite{raichlin2015protein} reconstructed using (\ref{fit}).
Start and end points of curves visually extracted from original figure. For activation energy the gap  between HOMO and HOMO-1 energies have been used in accordance with (\ref{tempidG}).\label{fig1}}
\end{figure} 
In Ref.\cite{raichlin2015protein} it has been found that the current through the system increases with the coupling strength if metallic electrodes are attached 
to various Myoglobin structures.   The change of the current with the temperature decreases with increasing strength and the current becomes 
temperature independent for small temperature. In all cases they found only mild temperature dependence inconsistent with the large HOMO-LUMO gap of Myoglobin. In Fig~\ref{fig1} we reproduced Figure 3. of the original article with
the formula 
\begin{equation}
    I=I_0+I_T\exp(-\Delta E/kT),\label{fit}
\end{equation}
where the parameter values are shown in Table~\ref{table1}. In Table~\ref{table2} we show the calculated values of the energies of molecular orbitals of Myoglobin. The  observed temperature dependence is not consistent with the large HOMO-LUMO gap and activation energy $\Delta E=(\varepsilon_{LUMO}-\varepsilon_{HOMO})/2=0.2226eV$, therefore we can exclude
all traditional explanations, which rely on the thermal excitation trough the gap. On the other hand (\ref{fit}) is fully consistent with
our formula (\ref{tempidG}) and (\ref{tempdpG}). According to our formula the conductance becomes
temperature independent for low temperatures and the weak temperature dependence is governed
by the smallest of the HOMO and HOMO-1 energy difference or LOMO and LOMO+1 energy difference. These 
are $\varepsilon_{LUMO+1}-\varepsilon_{LUMO}=0.6133 eV$ and  $\varepsilon_{HOMO}-\varepsilon_{HOMO-1}=0.0645eV$ for Myoglobin. The HOMO and HOMO-1
difference dominates and $\Delta E =\varepsilon_{HOMO}-\varepsilon_{HOMO-1}$ reproduces the experimental results correctly.
We note that in Myoglobin hole transport dominates the
temperature dependent part and that hole transport in general is often disregarded in intuition driven theoretical studies of electron transfer.
\begin{table}[htb]
\caption{Parameter values reproducing the Myoglobin measurement results of Ref.\cite{raichlin2015protein}.\label{table1}}
\centering
\begin{tabular}{lrr}
\toprule
&\textbf{$I_0$ in $A/cm^2$}	& \textbf{$I_T$ in $A/cm^2$} 	\\
\midrule
\text{Native m-Mb} &  $1.0 \cdot 10^{-7}$ 		& $3.2 \cdot 10^{-5}$		\\
\text{apo Mb}          & $1.0 \cdot 10^{-8}$    & $6.6 \cdot 10^{-6}$	 \\
\text{Reconstituted m-Mb}& $2.0 \cdot 10^{-6}$  & $2.7 \cdot 10^{-5}$\\
\bottomrule
\end{tabular}
\end{table}
\begin{table}[htb]
\caption{Myoglobin energies near the HOMO-LUMO gap. The energies have been calculated with the semiempirical extended Hückel method implemented in the YaEHMOP package \url{http://yaehmop.sourceforge.net}. Myoglobin structure taken from RCSB PDB \url{https://www.rcsb.org/structure/1MYF}.\label{table2}}
\centering
\begin{tabular}{lr}
\toprule
\textbf{}	& \textbf{energy in $eV$}	\\
\midrule
LUMO+1		& -8.9052		\\
LUMO 		& -9.5185	 \\
HOMO & -9.9637\\
HOMO-1& -10.0282\\
\bottomrule
\end{tabular}
\end{table}

\begin{figure}[htb]
\centering
\includegraphics[width=13 cm]{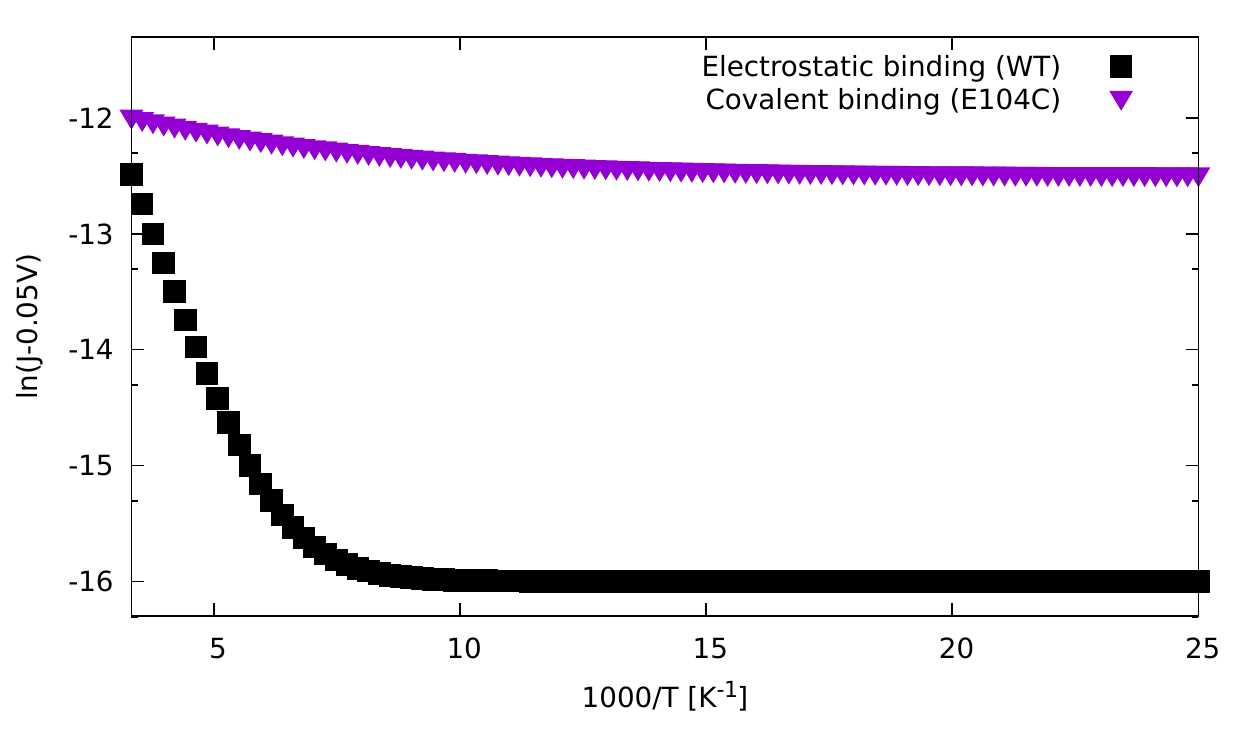}
\caption{Current density curves of experiment Ref\cite{amdursky2014solid} reconstructed using (\ref{fit}).
Start and end points of curves visually extracted from original figure. In case of covalent bonding for activation energy the gap  between LUMO and LUMO+1 energies have been used in accordance with (\ref{tempidG}), while for the electrostatic bonding case the experimentally found value has been used.\label{fig2}}
\end{figure}

Our second example is the measurement of Cytochrome C in Ref\cite{amdursky2014solid}. We reproduced two measurement curves shown in Figure 1.c. of the original paper and we show them in Fig~\ref{fig2}. 
We used again the fit (\ref{fit}) and the parameters are in Table~\ref{table3}. For the electrostatic case we used
the activation energy $\Delta E=0.105 eV$ found experimentally\cite{amdursky2014solid}. In Table~\ref{table4} we show
the calculated orbital energies of Cytochrome C. This experimentally found activation energy is in reasonable agreement with
the value calculated from the numerical  HOMO-LUMO gap $\Delta E=(\varepsilon_{LUMO}-\varepsilon_{HOMO})/2=0.161 eV$. According to our formula the conductance becomes
temperature independent for low temperatures and the weak temperature dependence is governed
by the smallest of the HOMO and HOMO-1 energy difference or LOMO and LOMO+1 energy difference, which 
are $\varepsilon_{LUMO+1}-\varepsilon_{LUMO}=0.0205 eV$ and  $\varepsilon_{HOMO}-\varepsilon_{HOMO-1}= 0.5336 eV$ for Cytochrome C. In this case the LUMO and LUMO+1
difference dominates and $\Delta E =\varepsilon_{LUMO+1}-\varepsilon_{LUMO}$ reproduces the experimental results correctly.
We note that in Cytochrome C electron transport dominates the temperature dependent part.
\begin{table}[htb]
\caption{Parameter values reproducing the Cytochrome C measurement results of Ref.\cite{amdursky2014solid}
.\label{table3}}
\centering
\begin{tabular}{lrrr}
\toprule
&\textbf{$I_0$ in $A/cm^2$}	& \textbf{$I_T$ in $A/cm^2$} & $\Delta E$ \textbf{in $eV$}	\\
\midrule
\text{Covalent binding (E104C)} &  $3.7 \cdot 10^{-6}$ 		& $5.3 \cdot 10^{-6}$ &	$0.020$	\\
\text{Electrostatic binding (WT)} & $1.1 \cdot 10^{-7}$ & $2.1 \cdot 10^{-4}$ & $0.105$\\
\bottomrule
\end{tabular}
\end{table}
\begin{table}[htb]
\caption{Cytochrome C energies near the HOMO-LUMO gap. The energies have been calculated with the semiempirical extended Hückel method implemented in the YaEHMOP package \url{http://yaehmop.sourceforge.net}. Cytochrome C structure taken from RCSB PDB \url{https://www.rcsb.org/structure/1HCR}.\label{table4}}
\centering
\begin{tabular}{lr}
\toprule
\textbf{}	& \textbf{energy in $eV$}	\\
\midrule
LUMO+1		&  -9.9457		\\
LUMO 		& -9.9252	 \\
HOMO & -9.6021\\
HOMO-1& -9.0685\\
\bottomrule
\end{tabular}
\end{table}

\subsection{Distance dependence of electron transfer and conductance}

In Ref.\cite{wierzbinski2013single} it has been found that in certain peptide nuclear acid structures both electron transfer rates and conductance decays exponentially with the length of the structure, but the two exponents differ considerably. While electron transfer rates decay as $e^{-\beta l}$, where 
$\beta_{ET} \approx 0.9 \mbox{\r{A}}^{-1}$ like in most biological structures, conductance decay is
about two thirds slower $\beta_G\approx 0.66\beta_{ET}$. The authors attributed this to a possible new
power law scaling relation between the two $G\sim k_{ET}^{0.66}$ in place of the linear relation
$G\sim k_{ET}$. As we have seen, there is no such a simple relation between these two quantities
in general, but it is possible that the two quantities decay with different exponents.

We note, that in the experiment the contact connecting the PNA structure to the Au electrode is weak.
This has been concluded in Ref.\cite{wolak2011electronic} 
suggesting that charge transfer between PNA and the Au substrate may be
difficult due to the required amounts of energy to overcome the injection barriers.
The metal electrode used in the conductance measurement has been strong indicated by the high values of measured conductance as well. Electron transfer rate in this situation is well  described by (\ref{ETtemp}) and conductance by (\ref{strongweakformula}) which can
be simplified further by keeping the leading term in case of a large HOMO-LUMO gap
\begin{equation}
G=\frac{e^2}{h}\left[ T^L_h(E_F)\frac{\Gamma^R_{HOMO}}{\Gamma^L_{HOMO}}+ T^L_e(E_F)\frac{\Gamma^R_{LUMO}}{\Gamma^L_{LUMO}}\right].
\end{equation}
The products of couplings show exponential dependence $\Gamma^{L}_{LUMO} \Gamma^{R}_{LUMO} \sim \Gamma^{L}_{HOMO} \Gamma^{R}_{HOMO}\sim e^{-\beta_{ET}l}$, where $l$ is the length of the molecule since they describe tunneling across the molecule. The HOMO and LUMO orbitals are somewhere
midway in the molecule and they are located very close to each other. It is then reasonable to assume that $\Gamma^{L}_{HOMO}\sim \Gamma^{L}_{LUMO}\sim e^{-\beta_{ET}xl}$ and $\Gamma^{R}_{HOMO}\sim \Gamma^{R}_{LUMO}\sim e^{-\beta_{ET}(1-x)l}$, where $xl$ and $(1-x)l$ is the distance of the left and right electrodes from the location of the HOMO-LUMO orbitals respectively. Inserting this into the
conductance and assuming that $T^L_h(E_F)$ and $T^L_e(E_F)$ don't change with the distance as we discussed before we can see that 
\begin{equation}
    G\sim e^{-\beta_{ET}(1-2x)l}.
\end{equation}
Assuming $x\approx 0.17$ can explain the relation of exponents observed in the experiment. This suggests that the HOMO and LUMO are closer to the metal contact, which  is due to the Ferrocene redox center attached to the end of the molecule in contact with the metal. This example clarifies that in case of a strong and a weak contact distance dependence can still be observed. Only two strong contacts guarantee distance independence.

\subsection{Distribution of conductance}

\begin{figure}[htb]
\centering
\includegraphics[width=13 cm]{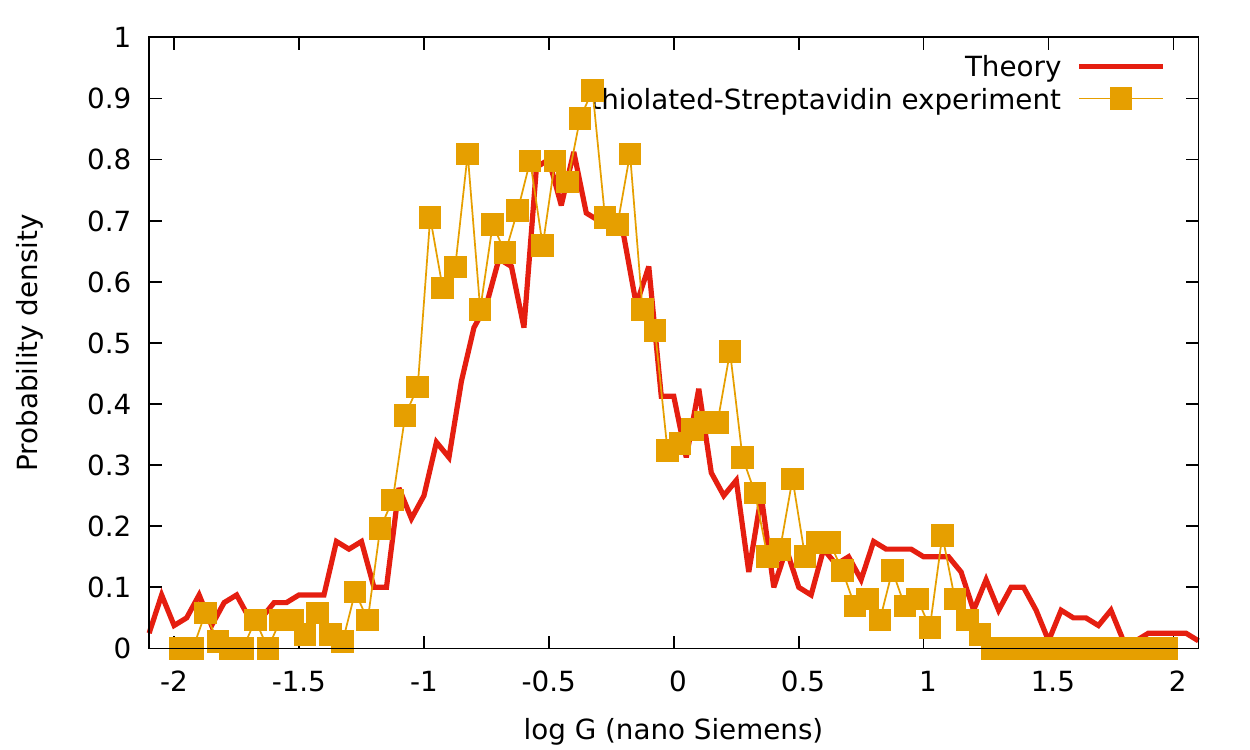}
\caption{Comparison of measurement data from Figure 4A of Ref.\cite{zhang2019electronic} (yellow points) and our
simulation (red line). Horizontal axis is logarithmic and conductance is in units of $nS$.\label{fig3}}
\end{figure}

In Ref.\cite{zhang2019electronic} the distribution of conductance between metallic contacts attached to various proteins has been investigated. Here we reconstruct the experiment where conductance have been measured between thiolated Lysines of a Streptavidin molecule. Streptavidin is the smallest protein used in Ref.\cite{zhang2019electronic} and it is
computationally feasible to calculate its electronic structure using semiempirical methods. For the calculations the 
structure \url{https://www.rcsb.org/structure/3RY1} has been used. Biotine molecules have been removed from the structure.
The energies and orbitals have been calculated with the semiempirical extended Hückel method implemented in the YaEHMOP package \url{http://yaehmop.sourceforge.net}. Here we attempt to reproduce the experimental result shown in Figure 2A (T-T),
in Figure 3A (lower part) and in Figure 4A of the original paper. Original data has been kindly provided by the Authors.
We assume that the thiolated sites have been randomly coupled to the substrate which and to the STM tip. We select pairs of thiolated sites using geometric information so that they lie on opposite sites of the molecule. We took all possible pairs with equal weight. For the coupling strengths the standard formula $\Gamma=V\mid \Phi_k^M \mid^2$ has been used, where $V$ is the strength of the coupling and $\mid \Phi_k^M  \mid^2$ is the quantum mechanical probability to find the electron in orbital $M$ on atom $k$. We assumed that the strengths are random and Gaussian distributed. We also assumed that the STM tip is strongly coupled with an average coupling strength $\overline{V}=0.4eV$ with
variance $0.35eV$, while we assumed a weaker coupling to the substrate  with average coupling strength $\overline{V}=0.07eV$ and variance $0.035eV$. These concrete values are the results of optimization carried out by visually comparing the result of the calculation with the experimentally obtained data. In Fig\ref{fig3} we show the result of this calculation.
The number of parameters in simulating the experimental situation is enormous and the details of the
structure of the protein and the quantum mechanical calculation contain a wide range of approximations and errors.
Never the less, it is obvious that (\ref{formula}) gives conductances in the correct order of magnitude with physically realistic coupling strengths. Both the average and the variance and the general shape of the resulting distribution is
compatible with the experimental finding. Note, that the inputs of (\ref{formula}) are the energies and the wave
functions of the molecule.
The theory presented here explains other aspects of this experiment such as the tip-substrate distance independence of the conductance distribution.

\section{Discussion}

The generalization of the Landauer-Büttiker formula revealed that decoherence plays an important
role in the electron transport properties of proteins and other biological molecules. Strong coupling
to vibration modes and the resulting decoherence explains the high conductance inside of these
structures. When accessed via weak electrostatic links these structures look like insulators since
electron transfer rate decaying with distance and temperature governs transport properties. When accessed
via strong covalent or nearly covalent bonds, the same structures show good conductance properties over
long distances and at high temperatures.  Beyond explaining novel experiments, what can be the biological 
role of these effects? In 1941 Nobel Prize winner biochemist Albert Szent-Györgyi put forward\cite{szent1941study} many examples when electrons travel over large distances very fast within a 
biomolecule or across the entire cell. Most of his problems are still open as Marcus theory strongly suppresses long distance transport over energy barriers. Strong decoherence via coupling of electronic
and vibrational degrees of freedom can open up new possibilities to understand the special electronic wiring
of biological systems.

\section{Materials and Methods}

Electronic properties of medium size proteins (Myoglobin, Cytochrome C and Streptavidin) have been 
calculated with the semiempirical extended Hückel method implemented in the YaEHMOP package \url{http://yaehmop.sourceforge.net}. Structures have been taken from RCSB PDB \url{https://www.rcsb.org}.
Computer codes for conductance calculations in Matlab and Phyton are provided as Supplementary 
material. 


\vspace{6pt} 



\section{Acknowledgements}
G.V. thanks Stuart Lindsay introduction and guidance in the subject and for sharing the data of his group's experiments.
This research was supported by the National Research Development and Innovation Office of Hungary (Project No. 2017-1.2.1-NKP-2017-00001).

\appendix
\section{}
We start with the definition of the inverse operator
\begin{equation}
   \sum_{pq}L_{nmpq}L^{-1}_{pqkl}=\delta_{nk}\delta_{ml}, 
\end{equation}
and separate the operator to a coupling strength dependent part plus the operator introduced in (\ref{L0}) 
\begin{equation}
    \sum_{pq}((-1/\hbar)(\Gamma_n/2+\Gamma_m/2)\delta_{np}\delta_{mq} + L_{nmpq}^0)L^{-1}_{pqkl}=\delta_{nk}\delta_{ml}.
\end{equation}
We take the diagonal elements ($n=m$) and get
\begin{equation}
    -\frac{\Gamma_n}{\hbar} L^{-1}_{nnkl}+ \sum_{pq}L_{nnpq}^0L^{-1}_{pqkl}=\delta_{nk}\delta_{nl}.
\end{equation}
We can summ up the left and right hand sides for $n$ and get
\begin{equation}
    -\sum_n\frac{\Gamma_n}{\hbar} L^{-1}_{nnkl} + \sum_{pq}\sum_n L_{nnpq}^0L^{-1}_{pqkl}=\delta_{kl}.
\end{equation}
Then $\sum_n L_{nnkl}^0=0$ due to the probability conservation law $\sum_nR_{nnkl}=0$, which can be be verified using (\ref{R}). 
Then setting $k=l$ we get the sum rule
\begin{equation}
    \sum_n\frac{\Gamma_n}{\hbar} L^{-1}_{nnkk}=-1. 
\end{equation}
The the expression (\ref{genLB}) of the current can be written also as
\begin{equation}
I=\frac{e^2U}{\hbar}\sum_n D_n(E_F,T) \left[\Gamma_n^L+\frac{1}{\hbar}\sum_k (\Gamma_k-\Gamma_k^R) L^{-1}_{kknn}(\Gamma_n^L-\Gamma_n^R)\right],
\end{equation}
and using the sum rule we can carry out the summation for $k$ for the first part and get
\begin{equation}
I=\frac{e^2U}{\hbar}\sum_n D_n(E_F,T) \left[\Gamma_n^R-\frac{1}{\hbar}\sum_k \Gamma_k^R L^{-1}_{kknn}(\Gamma_n^L-\Gamma_n^R)\right].\label{right}
\end{equation}
This expression is the same as (\ref{genLB}) just the $L$ and $R$ indices are interchanged. To get a formula manifestly symmetric in
these indices we have to add up (\ref{genLB}) and (\ref{right}) and divide by two to get (\ref{conductance}).

\section{}

We can start from the operator
\begin{equation}
    L_{nmkl}=(-1/\hbar)(\Gamma_n/2+\Gamma_m/2)\delta_{nk}\delta_{ml} + L_{nmkl}^0.
\end{equation}
The first term containing $\Gamma$ can be regarded as a perturbation to the second term.
The unperturbed system has a zero eigenvalue, since it corresponds to a closed system in
equilibrium
\begin{equation}
    \sum_{kl}L_{nmkl}^0\varrho_{kl}^0=0,
\end{equation}
where the right eigenvector is the equilibrium density matrix. Since electrons and holes are not mixed, we have two separate 
right eigenvectors $\varrho_{nn}^0=e^{-(\varepsilon_n-E_F)/kT}/Z(T)$ for
electrons (and zeros for holes) and $\varrho_{nn}^0=e^{-(E_F-\varepsilon_n)/kT}/Z(T)$ for holes (and zeros for electrons),
so there is an eigenvalue $\lambda=0$ both in the electron and in the hole sector. 
The left eigenvector
corresponding to the zero eigenvalue is $\delta_{nm}$ (assuming $n$ and $m$ are both electron or hole states), which we can verified by direct substitution
\begin{equation}
    \sum_{nm}\delta_{nm}L_{nmkl}^0=\sum_n L_{nnkl}^0=0,
\end{equation}
where we used the probability conservation law $\sum_nR_{nnkl}=0$, which can be be verified using (\ref{R}). In the first order of perturbation theory the new eigenvalue replacing the zero eigenvalue $\lambda=0$ becomes the perturbation
operator sandwiched between the left and the right unperturbed eigenvectors
\begin{equation}
    \lambda=\sum_{nmkl}\delta_{nm}(-1/\hbar)(\Gamma_n/2+\Gamma_m/2)\delta_{nk}\delta_{ml}\varrho_{kl}^0=-\sum_n \frac{\Gamma_n}{\hbar}\varrho_{nn}^0.
\end{equation}
The eigenvalues of the inverse operator $L_{nmkl}^{-1}$ are the reciprocals or the eigenvalues of the operator. Since the perturbed eigenvalue is very small, its reciprocal will dominate the 
inverse operator in leading order and we can write it in terms of the corresponding left and
right eigenvectors
\begin{equation}
    L_{nmkl}^{-1}\approx -\frac{\delta_{nm}\varrho_{nn}^0\delta_{kl}}{\sum_p \frac{\Gamma_p}{\hbar}\varrho_{pp}^0},
\end{equation}
where all four indices and the summation should be done for electron and hole states separately. 

\section{}

The left and right probabilities are dominated by the orbital close to the
Fermi energy. The first two most important contributions in case of holes 
come from the HOMO orbital and from HOMO-1 (the orbital below the HOMO)
\begin{equation}
    P^L_h=\frac{ \Gamma_{HOMO}^{L}e^{-(E_F-\varepsilon_{HOMO})/kT}+\Gamma_{HOMO-1}^{L}e^{-(E_F-\varepsilon_{HOMO-1})/kT}+...}{(\Gamma_{HOMO}^{L}+\Gamma_{HOMO}^R)e^{-(E_F-\varepsilon_{HOMO})/kT}+(\Gamma_{HOMO-1}^{L}+\Gamma_{HOMO-1}^R)e^{-(E_F-\varepsilon_{HOMO-1})/kT}+...}.
\end{equation}
This can be written as
\begin{equation}
    P^L_h=\frac{\Gamma_{HOMO}^{L}}{\Gamma_{HOMO}^{L}+\Gamma_{HOMO}^R}\frac{ 1+(\Gamma_{HOMO-1}^{L}/\Gamma_{HOMO}^{L})e^{-(\varepsilon_{HOMO}-\varepsilon_{HOMO-1})/kT}+...}{1+((\Gamma_{HOMO-1}^{L}+\Gamma_{HOMO-1}^R)/(\Gamma_{HOMO}^{L}+\Gamma_{HOMO}^R))e^{-(\varepsilon_{HOMO}-\varepsilon_{HOMO-1})/kT}+...},
\end{equation}
and then we can expand in the small parameter $e^{-(\varepsilon_{HOMO}-\varepsilon_{HOMO-1})/kT}$ and get in leading order and group
the terms
\begin{equation}
    P^L_h=\frac{\Gamma_{HOMO}^{L}}{\Gamma_{HOMO}^{L}+\Gamma_{HOMO}^R}+
    \frac{\Gamma_{HOMO}^{L}\Gamma_{HOMO}^{R}}{(\Gamma_{HOMO}^{L}+\Gamma_{HOMO}^R)^2}\left[\frac{\Gamma_{HOMO-1}^{L}}{\Gamma_{HOMO}^L}-   \frac{\Gamma_{HOMO-1}^{R}}{\Gamma_{HOMO}^R}     \right]e^{-(\varepsilon_{HOMO}-\varepsilon_{HOMO-1})/kT}+... \:  .
\end{equation}
The sign of the temperature dependent part is determined by the ratios $\Gamma_{HOMO-1}^{L}/\Gamma_{HOMO}^L$ and $\Gamma_{HOMO-1}^{R}/\Gamma_{HOMO}^R$ which are the relative strengths of couplings of the left and right electrodes
to the HOMO and HOMO-1 orbitals. By exchanging left and right we can get the probability
\begin{equation}
    P^R_h=\frac{\Gamma_{HOMO}^{R}}{\Gamma_{HOMO}^{L}+\Gamma_{HOMO}^R}-
    \frac{\Gamma_{HOMO}^{L}\Gamma_{HOMO}^{R}}{(\Gamma_{HOMO}^{L}+\Gamma_{HOMO}^R)^2}\left[\frac{\Gamma_{HOMO-1}^{L}}{\Gamma_{HOMO}^L}-   \frac{\Gamma_{HOMO-1}^{R}}{\Gamma_{HOMO}^R}     \right]e^{-(\varepsilon_{HOMO}-\varepsilon_{HOMO-1})/kT}+... \: .
\end{equation}
In case of electrons the LUMO and LUMO+1 orbitals play the same role and we can get the analogous expression
\begin{equation}
    P^L_e=\frac{\Gamma_{LUMO}^{L}}{\Gamma_{LUMO}^{L}+\Gamma_{LUMO}^R}+
    \frac{\Gamma_{LUMO}^{L}\Gamma_{LUMO}^{R}}{(\Gamma_{LUMO}^{L}+\Gamma_{LUMO}^R)^2}\left[\frac{\Gamma_{LUMO+1}^{L}}{\Gamma_{LUMO}^L}-   \frac{\Gamma_{LUMO+1}^{R}}{\Gamma_{LUMO}^R}     \right]e^{-(\varepsilon_{LUMO+1}-\varepsilon_{LUMO})/kT}+...,
\end{equation}
and
\begin{equation}
    P^R_e=\frac{\Gamma_{LUMO}^{R}}{\Gamma_{LUMO}^{L}+\Gamma_{LUMO}^R}-
    \frac{\Gamma_{LUMO}^{L}\Gamma_{LUMO}^{R}}{(\Gamma_{LUMO}^{L}+\Gamma_{LUMO}^R)^2}\left[\frac{\Gamma_{LUMO+1}^{L}}{\Gamma_{LUMO}^L}-   \frac{\Gamma_{LUMO+1}^{R}}{\Gamma_{LUMO}^R}     \right]e^{-(\varepsilon_{LUMO+1}-\varepsilon_{LUMO})/kT}+... \: .
\end{equation}

\section{}

We start with the definition of the inverse operator
\begin{equation}
   \sum_{pq}L^{-1}_{nmpq}L_{pqkl}=\delta_{nk}\delta_{ml}, 
\end{equation}
and separate the operator to a coupling strength dependent part plus the operator introduced in (\ref{L0}) 
\begin{equation}
    \sum_{pq}L^{-1}_{nmpq}((-1/\hbar)(\Gamma_k/2+\Gamma_l/2)\delta_{pk}\delta_{ql} + L_{pqkl}^0)=\delta_{nk}\delta_{ml}.
\end{equation}
We take the diagonal elements ($k=l$) and get
\begin{equation}
    -L^{-1}_{nmkk}\frac{\Gamma_k}{\hbar} + \sum_{pq}L^{-1}_{nmpq}L_{pqkk}^0=\delta_{nk}\delta_{mk}.
\end{equation}
We can multiply both sides with the Boltzmann distribution $p_k^B$, sum up for $k$ and get
\begin{equation}
    -\sum_k L^{-1}_{nmkk}\frac{\Gamma_k}{\hbar}p_k^B + \sum_{pq}L^{-1}_{nmpq}\sum_k L_{pqkk}p_k^B=\delta_{nm}p_n^B
\end{equation}
Then $\sum_k L_{pqkk}^0p_k^B=0$ since the Boltzmann distribution is the steady state solution. 
Then setting $m=n$ we get the sum rule
\begin{equation}
    \sum_k L^{-1}_{nmkk}\frac{\Gamma_k}{\hbar}p_k^B=-p_n^B.
\end{equation}

\end{document}